\documentstyle[preprint,tighten,aps]{revtex}

\newcommand{\be}{\begin{equation}}
\newcommand{\ee}{\end{equation}}
\newcommand{\bea}{\begin{eqnarray}}
\newcommand{\beas}{\begin{eqnarray*}}
\newcommand{\eea}{\end{eqnarray}}
\newcommand{\eeas}{\end{eqnarray*}} 
\newcommand{\ba}{\begin{array}}
\newcommand{\ea}{\end{array}}

\begin{document}

\vspace*{2.5cm}
{\Large\bf Minimal Coupling and Feynman's Proof}

{\bf Merced Montesinos\footnote{
        Department of Physics and Astronomy,
        University of Pittsburgh, Pittsburgh PA 15260, USA.}
        \footnote{ 
        Departamento de F\'{\i}sica, Centro de Investigaci\'on y de
        Estudios Avanzados del I.P.N.,\\
        Av. I.P.N. No. 2508, 07000 Ciudad de M\'exico, M\'exico.
        E--mail: merced@fis.cinvestav.mx} and
        Abdel P\'erez-Lorenzana\footnote{
        Departamento de F\'{\i}sica, Centro de Investigaci\'on y de
        Estudios Avanzados del I.P.N.,\\
        Av. I.P.N. No. 2508, 07000 Ciudad de M\'exico, M\'exico.
        E--mail: abdel@fis.cinvestav.mx}}

\begin{center}
\begin{minipage}[t]{9.5cm}\footnotesize
\hrulefill

The non quantum relativistic version of the proof of Feynman
for the Maxwell equations is discussed in a
framework with a minimum number of hypotheses required. From the present
point of view it is clear that the classical equations of motion 
corresponding to the gauge field interactions can be deduced from the 
minimal coupling rule, and we claim here resides the essence of the proof 
of Feynman.

\hrulefill
\end{minipage}
\end{center}

\section{Introduction}
The proof of Feynman for the Maxwell equations presented in Dyson's 
paper (Dyson,1990) was never published by Feynman himself 
because from his point of view the proof provides no new information about 
the classical or quantum nature of the electromagnetic field. Even though the 
proof is mathematically right, there are mixing physical inputs. First
of all, the proof is based on the second law of Newton, which is a classical 
relation. Second, the quantum commutators between position and 
momentum are assumed (Dombey, 1991; Brehme, 1991; Anderson, 1991; 
Farquhar, 1991). Third, the framework is Galilean. Therefore, is quite 
surprising that with this information pure relativistic equations of motion 
emerge from this formalism. This is in fact the main result of the proof of 
Feynman. It is important to emphasize that the proof reproduces only the 
homogeneous Maxwell equations, which in fact are compatible with Galilean 
relativity (Vaidya {\it et al}., 1991).

On the other hand, Dyson claims that the proof has a remarkable property in 
that it shows that the physics involved in the assumptions concerns only 
the homogeneous Maxwell equations. There are some results which show that 
Feynman's proof can be extended to the non-Abelian gauge fields (Lee, 1990) 
within a relativistic framework (Tanimura, 1992). Nevertheless, these last
proposals have the same mixed physical inputs as the original proof
(Farquhar, 1990). A straightforward extension was reviewed recently
by Bracken (1998) who gave a derivation of 
the homogeneous Maxwell equations from a postulated set of Poisson 
brackets instead of the quantum commutators (just like 
Hughes, 1992). Bracken also proposed an extended formalism by postulating a 
set of relativistic Poisson brackets. Nevertheless this approach is, on one 
hand, non manifestly covariant, and on the other hand, it is unable to 
derive the nonhomogeneous Maxwell equations, even though the field tensor 
may be built.

At this point, it seems that the most important physical property
associated with the dynamics of a particle under the action of a 
gauge field is missed in all this approaches, the minimal coupling
rule. It contains all the information of the sources and fields,
and therefore it is a more natural starting point. In fact, it is
equivalent to the assumptions involved in Feynman's proof, but
it has the advantage of having a clear physical meaning; this is our 
main claim. Nothing of this seems to be new; however, this proof 
has attracted interest in the community because of its relationship 
with some fundamental aspects of physics. Thus our main motivation 
for giving the proof again is to illuminate its physical basis.   

In keeping with this goal, our proof uses the minimum of hypotheses. It is 
based on the assumption that the minimal coupling rule holds. No quantum
commutation relations are assumed. Section II reviews the original
proof, Section III shows that Feynman's hypothesis can be obtained from the 
minimal coupling rule for a relativistic particle. So from the perspective 
of the present approach the validity of the minimal coupling rule is the 
essential element underlying Feynman's construction. Using the results of 
this section, we exhibit our approach explicitly in Sections. IV and
V, where the electromagnetic and the non-Abelian fields are 
considered, respectively.

\section{The proof of Feynman}
We begin by reviewing the Feynman's proof. Essentially we follow the
same approach as in Dyson (Dyson, 1990); our notations and conventions 
are the same. Let us consider a free particle with position and velocity 
$x_i$, and $\dot x_i= dx_i/dt$, respectively. Then Newton's second law holds:
\be
m\ddot x_j = F_i(x,\dot x,t) \label{1}\, .
\ee
Also the quantum commutation relations are assumed:
 \be   [x_j,x_k]=0, 
\qquad \qquad  m[x_j,\dot x_k] = i\hbar \delta_{jk}\ . 
\label{3}\ee
Then (\ref{1}) and (\ref{3}) imply 
 \be [x_j,F_k] + m[\dot x_j,\dot x_k] = 0. \label{9} \ee
Now because $[x_j,F_k]$ is skew symmetric in the pair $j$ and $k$, it
allows us to introduce the auxiliary field $H_l$ through
 \be [x_j,F_k] = -{i\hbar\over m}\epsilon_{jkl}H_l\, ,
\label{13}\ee
and by using the Jacobi identity 
$[x_l,[\dot x_j,\dot x_k]] + [\dot x_j,[\dot x_k,x_l] + 
[\dot x_k,[x_l,\dot x_j]] = 0$ 
together with (\ref{3}) and (\ref{9}) is straightforward to see
$H_l$  depends only on $x$ and $t$ because
$[x_l,[x_j,F_k]]=0$, 
or which is the same
 \be [x_l,H_m] = 0.
\label{14} \ee
It is convenient to define a new field $E_j$ by 
employing the relation
 \be F_j = E_j + \epsilon_{jkl}\,\dot x_k H_l, 
\label{4}\ee
which from (\ref{3}), (\ref{13}), and (\ref{14}) satisfies 
$[x_m,E_j]=0$, which means it does not depend on $\dot x$.
On the other hand, by using (\ref{9}) and (\ref{13}), we can get for 
$H_l$ 
 \be H_l = {m^2\over i2\hbar}\epsilon_{jkl} [\dot x_j, \dot x_k].
 \label{16} \ee
which together with the Jacobi identity allows us to obtain
 \be {\partial H_l\over\partial x_l} = 
[\dot x_l,H_l] = {m^2\over i2\hbar}
\epsilon_{jkl}[\dot x_l,[\dot x_k,\dot x_j]] = 0. \label{17} 
\ee
Next we take the total derivative of (\ref{16}) with respect to $t$ 
 \be {\partial H_l \over \partial t} + \dot x_m {\partial H_l \over
\partial x_m} = {m^2\over i\hbar} \epsilon_{jkl}[\ddot x_j,\dot x_k]. 
\label{19} 
\ee
Finally, from (\ref{1}) and (\ref{4}) the RHS of (\ref{19}) 
can be written as
 \bea 
{m\over i\hbar}\epsilon_{jkl}[E_j + \epsilon_{jmn}\dot x_m H_n, \dot
x_k]
&=& {m\over i\hbar}\big( \epsilon_{jkl}[E_j,\dot x_k] + 
[\dot x_k H_l, \dot x_k]
-[\dot x_l H_k,\dot x_k] \big)\nonumber \\
&=& \epsilon_{jkl} {\partial E_j \over \partial x_k} +
\dot x_k{\partial H_l \over \partial x_k} +
\dot x_l{\partial H_k \over \partial x_k} +
 {m\over i\hbar}H_k[\dot x_l,\dot x_k]. \label{20} 
\eea
In the last expression the third and fourth terms vanish because of
(\ref{16}) and (\ref{17}). Therefore, by putting (\ref{20}) 
in (\ref{19}), we obtain Faraday's induction law,
 \be
{\partial H_l\over \partial t}  =
\epsilon_{jkl}{\partial E_j\over \partial x_k}.
\label{21} 
\ee
End of the proof. 

Now, here it is important to make some remarks. First, note that the 
Galilean version of the Lorentz Law, Eq. (\ref{4}), has been 
explicitly used. Moreover, (\ref{3}) means we are using a quantum 
framework. In other words, there are two mixed inputs: classical and 
quantum descriptions are combined in Feynman's proof. Even
though these two classical and quantum aspects are taken into account,
the result is amazing. The main result of the Feynman's proof is 
that, from quantum commutators [Eq. (\ref{3})] and the quantum version
of the Newtonian force (Tanimura, 1992), the equations of motion for
the fields are the homogeneous Maxwell equations. The proof can be
extended to the case of non-Abelian gauge fields both in Newtonian
(Lee, 1990) as well as in relativistic (Tanimura, 1992) dynamics. It
is clear that because of the nature of the fields only the
relativistic approach allows the construction of all equations of
motion for the fields.  Note also that Feynman's proof requires
only the quantum commutation relations (Hughes, 1992), which for the
purposes of the proof can be substituted by their classical version, the 
Poisson brackets. This key property raises the possibility of constructing 
a nonquantum version of the proof in the framework of special relativity 
with a minimum of hypotheses: the minimal coupling rule.

\section{Relativistic case}

Let us consider a relativistic particle, with rest mass
$m$, in a inertial frame under the action of an external force
in such a way that the generalized momentum satisfies the minimal 
coupling rule (see, for instance, O'Raifeartaigh, 1997)
 \be
\pi_\mu = m \dot x_\mu + A_\mu (x,\pi), \label{am}
\ee
where $\pi_\mu$ is the canonical momentum of the particle, which has the 
contribution of the fields through the potential $A_\mu$. In a general 
situation $A_{\mu}$ might depend in the velocity of the particle 
or, which is the same, on the components of the canonical 
momentum. Consequently, the physical gauge fields will be deduced from 
particular restrictions on this dependence, as we shall show in next 
section. In fact, the proof we present below is totally general. 

From now on, we will denote the derivative with respect to the
proper time $\tau$ as well as the derivatives with respect to 
the canonical coordinates of a phase space function $f(x,\pi)$ by
\be
\dot f \equiv {df\over d\tau}, \qquad  
\partial^\mu \equiv {\partial \over \partial x_\mu},\qquad 
\bar \partial^\mu \equiv {\partial \over \partial \pi_\mu}\, .
\ee
In this way, the Poisson bracket is given by
 \be
\{f,g\}\equiv \eta_{\rho\sigma}(\partial^{\rho}f \bar
\partial^{\sigma} g - \partial^{\rho}g \bar \partial^{\sigma}f) ,
\label{uno}
\ee
where $\eta_{\rho\sigma}$ is the Minkowski metric.

Instead of taking Feynman's hypothesis, we are going to assume
that the relation (\ref{am}) holds. In other words, the relation 
(\ref{am}) is put on a fundamental level, and using it, we shall show 
that the equations of motion for the fields, and the interaction
law with a test particle can be deduced without any additional
assumption. So the present approach shows the minimal coupling rule
has itself all the dynamic information of the system.

From the definition of the Poisson brackets and the minimal coupling
rule we get the relationship
 \be
m \{ x_\mu , \dot x_\nu \} = \eta_{\mu\nu} - \bar \partial_\mu A_\nu ,
\label{px}
\ee
which is the analog of (\ref{3}) in the nonrelativistic case. Taking the 
derivative of (\ref{px}) with respect to the proper time 
$\tau$, we have
 \be
m {d \over d\tau} \{ x_\mu , \dot x_\nu \} =  
m \{ \dot x_\mu , \dot x_\nu \} + m \{ x_\mu,\ddot x_\nu \} =  
- {d \over d\tau} (\bar \partial_\mu A_\nu ).  \label{dpx}
\ee
Now, using the Jacobi identity for
$\{ x_\nu ,\{\dot x_\mu,\dot x_\rho \} \}$ and (\ref{px}), one obtains
 \be
m \{ x_\nu , \{ \dot x_\mu , \dot x_\rho \} \} +
\{ \dot x_\mu , \bar \partial_\nu A_\rho \} +
\{ \bar \partial_\nu A_\mu , \dot x_\rho \} = 0,
\ee
which means that the quantity $\{ \dot x_\mu ,\dot x_\rho \}$ depends on
the  derivative of $x$, through the implicit dependence of $A_\mu$ on
$\pi_\mu$. This is the most general situation [compare  with Eq.
(\ref{14})]. Following Feynman, we define the skew symmetric tensor 
 \be
- {1 \over m} F_{\mu\nu} \equiv - m \{ \dot x_\mu , \dot 
x_\nu \} = m  \{ x_\mu , 
\ddot x_\nu \} + {d \over d\tau} ( \bar \partial_\mu A_\nu ),
 \label{fmnx}
\ee 
which, after it is expanded, takes the form of the tensor associated
to the gauge field $A_\mu$ 
 \be  
 F_{\mu\nu} = ( \partial_\mu A_\nu - \partial_\nu A_\mu ) + 
\{ A_\mu , A_\nu \} \,. \label{fmn}
\ee
Notice that the last term of (\ref{fmn}) must vanish for the 
electromagnetic case (see next section). In general it suggests the right 
form of the non-Abelian gauge field tensor. Taking the derivative with 
respect to $x_\alpha$, we obtain the following relation:
\be
\partial_\alpha F_{\mu\nu} + \partial_\mu F_{\nu\alpha} 
+ \partial_\nu 
F_{\alpha\mu} = \partial_\alpha \{ A_\mu , A_\nu \} + 
\partial_\mu \{ A_\nu , 
A_\alpha \} + \partial_\nu \{ A_\alpha , A_\mu \}, \label{cicli}
\ee
which suggests the definition of a ``covariant derivative'' of the
form 
$D_\alpha F_{\mu\nu} \equiv \partial_\alpha F_{\mu\nu} -
\left\{F_{\mu\nu}, A_\alpha\right\}$. 
This implies (\ref{cicli}) can be written as 
 \be
D_\alpha F_{\mu\nu} +D_\mu F_{\nu\alpha} + D_\nu F_{\alpha\mu} = 0.
\label{bianchi}
\ee
The former expression corresponds in general to the homogeneous field 
equations. This identity is in fact equivalent to the equations obtained 
from the usual approaches (Dyson, 1990; Lee, 1990; Hughes, 1992; 
Tanimura, 1992; Bracken, 1998). Now, since 
$\partial^{\mu}\partial^{\nu} F_{\mu\nu} = 0$ holds, there must must exist 
a conserved current given by 
\be 
j_\mu \equiv \partial^{\nu} F_{\mu\nu}, \label{jmu} 
\ee
which as usual can be identified as the source of the fields
(Jackson, 1975). Therefore, the last equation corresponds to the 
nonhomogeneous field equation which is not obtained in the original
scheme by Feynman (Dyson, 1990), nor in the extended versions of the proof
(Lee, 1990; Hughes, 1992; Tanimura, 1992; Bracken, 1998). This is the most 
relevant equation, for it defines the dynamics of the fields
(Jackson, 1990). Now, starting from (\ref{fmnx}), which 
defines $F_{\mu\nu}$, we note that the relation 
 \be 
F_{\mu\nu}\dot x^\nu = 
\{m\dot x_\mu,{1\over 2}m\dot x_\nu \dot x^\nu\},
\ee
holds, which suggests including the Hamiltonian of the system 
(Goldstein, 1980)
 \be
 H = {1\over 2m}\left(\pi - A\right)^2 = {1\over 2}m\dot x_\nu \dot
x^\nu, \label{H}
\ee
and obtaining a generalized Lorentz law
\be
F_{\mu\nu}\dot x^\nu = m\ddot x_\mu.\label{fl}
\ee
In summary, starting only from the minimal coupling rule (\ref{am}), we 
were able to obtain  the tensor of the interaction fields
$F_{\mu\nu}$ as well as the equation of motion of a test particle
(\ref{fl}) and the analog to the field equations [Eqs.
(\ref{bianchi}) and (\ref{jmu})]. In the next two sections we will apply
explicitly this method to both the Abelian and non-Abelian cases. 

\section{The Abelian case: electromagnetic field }

The electromagnetic case is the simplest one. Let us take the
following restriction on the fundamental hypothesis [given by
Eq. (\ref{am})]:
 \be
A_\mu = A_\mu(x) , \label{amem}
\ee
i.e., $\bar \partial_\mu A_\nu = 0$, which means, for the
present case, (\ref{px}) reduces to $m \{ x_\mu , \dot x_\nu \} =
\eta_{\mu\nu}$ (the usual starting point of Feynman's proof). Note also 
that $\{ A_\mu , A_\nu \} = 0$ holds. Therefore, (\ref{fmn}) allows us to 
define the electromagnetic tensor
 \be
F_{\mu\nu} (x) = \partial_\mu A_\nu - \partial_\nu A_\mu ,
\label{fem} 
\ee
which satisfies the Bianchi identity [obtained from (\ref{cicli})] 
 \be
\partial_\mu F_{\nu\alpha} + \partial_\nu F_{\alpha\mu} 
+ \partial_\alpha  F_{\mu\nu } = 0\, ,\label{cicliem}
\ee
which corresponds to the homogeneous Maxwell equations. The equations 
with sources can be gotten from
$j_\mu \equiv \partial^{\nu} F_{\mu\nu}$ in (\ref{jmu}). Explicitly, as 
usual, $E_i = F_{0i}$ and $H_i = {\tilde F}_{0i}$, where
${\tilde F}_{\mu\nu} = 
{1\over 2} \epsilon_{\mu\nu\alpha\beta}F^{\alpha\beta}$ 
is the dual tensor, and $i=1,2,3$. Consequently, the electromagnetic 
fields are defined by
${\bf E} = -\partial_0 {\bf A}- \nabla A_0$ and 
${\bf H} =  \nabla\times {\bf A}$. 

Note also that the Lorentz law is clearly (\ref{fl}). We have used 
explicitly the Hamiltonian of the test particle in order to obtain this 
equation of motion, but this expression can be obtained by integrating
(\ref{fmnx}) with respect to $\pi_\mu$. 

In summary, we have obtained the complete set of the Maxwell equations
and the Lorentz law for the  test particle (without assuming it from
the beginning) just starting from the hypothesis (\ref{amem}). It is
important to emphasize this  property, of the present approach, because 
it is not shared by the Feynman's proof (Dyson, 1990) nor its direct 
extensions (Lee, 1990; Hughes, 1992; Tanimura, 1992; Bracken, 1998). In 
others words, the four Maxwell equations (with sources and without 
magnetic monopole terms) emerge in a natural way if a coupling of the form 
(\ref{amem}) is assumed, which means that $A_\mu$ does not depend
on the  velocity of the test particle. Therefore, all the dynamic 
information of the system is contained in the minimal coupling rule, as we 
claimed.

\section{The non-Abelian gauge fields}

As we noted above, the general form of the field tensor [Eq. 
(\ref{fmn})] is that of the non-Abelian case. It suggests the
classical non-Abelian gauge field equations could be obtained from
the minimal coupling rule through a special condition over $\{ A_\mu
, A_\nu \}$. Basically we have to note that in the classical approach
the non-Abelian fields may be treated by introducing new internal
degrees of freedom, as the isospin, in such a way that the
Hamiltonian would depend in some other non spacetime
variables. Some steps in this direction were made by Lee (1990) and 
Tanimura (1992) and recently discussed in a relativistic context 
by Bracken (1998).

Let us consider as the canonical coordinates of the test particle 
those which belong to a $d+n$ dimensional space, where $d$ is the
spacetime dimension, and $n$ is the  internal space dimension (for
instance, isospin), which is necessary to ``balance'' the momentum
due to the external interaction. We use the following notations and
conventions, $\Omega,\Lambda = 0,1\dots d\dots d+n$; $\alpha,
\mu,\nu = 0,1,\dots d$; and  $a,b,c=d+1,\dots, d+n$; in such a way
that all the results obtained in the  section III hold on the indices
$\Omega,\Lambda$.

Next, let us assume the dependence on the canonical coordinates
of  $A_\Omega$ is such that it is separable, and can be written in the form
 \be
A_\Omega = A_\Omega (x_\Lambda , \pi_a ) = A_{\Omega a}(x_\mu) I^a 
(x_b,\pi_b), \label{ana}
\ee 
which means it does not depend on the canonical momentum associated
to the spacetime coordinates. Also let us assume that $I^a$ satisfies
 \be
\{ I^a , I^b \} = - {f^{ab}}_c I^c ,\label{grup}
\ee
where ${f^{ab}}_c$ are the structure constants corresponding to the
Lie algebra of the Lie group locally generated by the quantum 
operators associated to the functions $I^a$. 

Note that due to the separation of the coordinates, the Poisson
brackets can be written in the form
$\{ A , B \} = \{ A , B \}_{esp} + \{ A , B \}_{int}$
where ``esp'' and ``int'' mean spacetime and internal 
space, respectively. Under this consideration and taking (\ref{ana})
into  account, we will  have
$\{ A_\Omega , A_\Lambda \} =  A_{\Omega a} 
A_{\Lambda b} \{ I^a , I^b \}$, for the spacetime part of the
bracket vanishes. Hence,  expressing  the equations only in the 
spacetime coordinate sector we find that (\ref{fmn}) can be written
as $F_{\mu\nu} = 
\left( \partial_\mu A_{\nu c} - \partial_\nu A_{\mu c} - 
A_{\mu a} A_{\nu b} {f^{ab}}_c \right) I^c $,      
from which it is natural to interpret the term between brackets
in the former equation as the {\it Yang-Mills field} tensor given
by
\be
F_{\mu\nu c} \equiv \partial_\mu A_{\nu c} - \partial_\nu A_{\mu c} -
A_{\mu a} A_{\nu b} {f^{ab}}_c . \label{ym}
\ee
Note that because $A_\Omega$ does not depend on $\pi_\nu$, the
equation of motion (\ref{fl}) can also be obtained by integrating
(\ref{fmnx}), acquiring the form
\be
m \ddot x_\mu = F_{\mu\nu a} I^a \dot x^\nu  + G_{\mu a} 
I^a , \label{1wong}
\ee
which is the {\it first Wong equation for the non-Abelian gauge
fields}. As we shall see, the last term in the equation above, absent
from (\ref{fl}), may be identified as a gauge term (see last part of
this section).

On the other hand, the covariant derivative can be defined as
 \be
(D_\alpha F_{\mu\nu} )_c \equiv \partial_\alpha F_{\mu\nu c} - 
{f^{ba}}_c  A_{\alpha b} F_{\mu\nu a}, \label{dc}
\ee
implying the Bianchi identity
\be
( D_\alpha F_{\mu\nu} )_c + ( D_\mu F_{\nu\alpha} )_c 
+ ( D_\nu F_{\alpha\mu} )_c = 0 . \label{ciclina}
\ee
Also from
$\dot I^a = \{I^a,H\} = m\dot x^\mu\{I^a,\dot x_\mu\}$
and 
\be m \{I^a,\dot x_\mu\} = A_{\mu b} I^c {f^{ab}}_c \label{ib}\ee
we can get the {\it second Wong equation}
\be
 \dot I^a - {f^{ab}}_c A_{\nu b} \dot x^\nu I^c = 0, \label{2wong}
\ee
or  equivalently 
$\{ x_\mu , ( \dot I^a - {f^{ab}}_c A_{\nu b} \dot x^\nu I^c) \}=0$.
Finally, from (\ref{ib}) we obtain the usual expression for 
functions of the type $\phi_a (x)$:
 \be
m \{ \dot x_\mu , \phi_a (x) I^a \} =
 - \{ \partial_\mu \phi_a 
 - {f^{bc}}_a A_{\mu b} \phi_c \} I^a =
 -  (D_\mu \phi)_a I^a .
\ee
In particular $G_\mu \equiv G_{\mu a} I^a$ satisfies
$m \{ \dot x_\mu , G_\nu \} = -( D_\mu G_\nu)_a I^a$,
which together with the two Wong equations implies $G_{\mu}$ is a
gauge  term because
 \be ( D_\mu G_\nu )_a - ( D_\nu G_\mu )_a = 0,   \ee
holds.
 
\section{Concluding remarks}

We summarize the above results as follows. Even though Feynman's proof 
fails because it provides no new physics, the proof is
successful because it reduces the laws of the gauge interactions in
the sense that they can be obtained from only the minimal coupling
postulate. This fact is not only an economic choice, but it has a 
deeper meaning which for the present analysis signifies that the
fundamental dynamic equations, the field equations, and the motion
equation of the test particle  (the Lorentz law) just come from
the minimal coupling rule between the potential and the linear
momentum. However, this fundamental fact is unclear (and missed in
the discussions) of the approaches that start from postulating the
quantum commutators or the equivalent Poisson brackets.

On the other hand, it is important to emphasize some aspects of the 
present approach. First, we are in the framework of relativistic
classical mechanics. Second, a quantum point of view has not been
adopted, so the relationship with the quantization schemes should 
be analyzed carefully. In particular, the relationship with the  Dirac
method (Dirac, 1964; Henneaux and Teitelboim, 1992) would be interesting 
of studying because of the implicit dependence of the gauge fields on 
the momenta [see (\ref{am})]. We are aware the present approach might 
be introducing second-class constraints in a quantum analysis (as 
in QED). If this were the case, a more careful treatment should be given if a 
quantization based on the present approach is considered. These 
conjectures are beyond the scope of the present paper, but they should 
be clarified.

\section*{Acknowledgments}

M.M. thanks all the members of the relativity group of the Department
of Physics and Astronomy of the University of Pittsburgh for their
warm  hospitality. M.M.'s postdoctoral fellowship is funded through 
the CONACyT of Mexico, Fellow No. 91825. The authors also acknowledge the 
support of the {\it Sistema Nacional de Investigadores} of CONACyT.


\end{document}